\documentclass[sigconf,noncam]{acmart}

\usepackage{algorithm}
\usepackage{algorithmicx}
\usepackage[noend]{algpseudocode}
\usepackage{bm}
\usepackage{subcaption}
\DeclareMathOperator*{\argmin}{argmin}

\usepackage{balance}
\usepackage{multirow}

\AtBeginDocument{%
  }

\copyrightyear{2023} 
\acmYear{2023} 
\setcopyright{acmlicensed}\acmConference[MM '23]{Proceedings of the 31st ACM International Conference on Multimedia}{October 29-November 3, 2023}{Ottawa, ON, Canada}
\acmBooktitle{Proceedings of the 31st ACM International Conference on Multimedia (MM '23), October 29-November 3, 2023, Ottawa, ON, Canada}
\acmPrice{15.00}
\acmDOI{10.1145/3581783.3612290}
\acmISBN{979-8-4007-0108-5/23/10}

\renewcommand\footnotetextcopyrightpermission[1]{}
\begin{document}

\title{Relative NN-Descent: A Fast Index Construction for Graph-Based Approximate Nearest Neighbor Search}

\author{Naoki Ono}
\email{n\_ono@hal.t.u-tokyo.ac.jp}
\orcid{0009-0005-7358-7070}
\affiliation{%
  \institution{The University of Tokyo}
  \city{Tokyo}
  \country{Japan}
}

\author{Yusuke Matsui}
\email{matsui@hal.t.u-tokyo.ac.jp}
\orcid{0000-0003-1529-0154}
\affiliation{%
  \institution{The University of Tokyo}
  \city{Tokyo}
  \country{Japan}
}

\renewcommand{\shortauthors}{Naoki Ono \& Yusuke Matsui}

\begin{abstract}
  % 近似最近傍探索（ANNS）は，与えられたクエリベクトルに最も近いベクトルをデータベースから探すタスクである．グラフベースのANNSは，million-scaleのデータセットに対して最も精度・速度のバランスが良い手法の系列である．一方で，グラフベースの手法はインデックスの構築時間が長いという弱点を持つ．近年多くの手法が，探索時の精度と速度のトレードオフを改善している．しかし，インデックスの構築を高速化する研究はほとんど存在しない．
  Approximate Nearest Neighbor Search (ANNS) is the task of finding the database vector that is closest to a given query vector. Graph-based ANNS is the family of methods with the best balance of accuracy and speed for million-scale datasets. However, graph-based methods have the disadvantage of long index construction time. Recently, many researchers have improved the tradeoff between accuracy and speed during a search. However, there is little research on accelerating index construction.
  % 我々は高速なグラフ構築アルゴリズムであるRalative NN-Descent (RNN-Descent) を提案する．RNN-Descentは，近似K近傍グラフ (K-NN) を構築するアルゴリズムであるNN-Descentと，探索に有効な辺を選択するアルゴリズムであるRNG Strategyを組み合わせたものである．このアルゴリズムはグラフインデックスを，ANNSに基づかない方法で直接構築することを可能にする．
  We propose a fast graph construction algorithm, Relative NN-Descent (RNN-Descent). RNN-Descent combines NN-Descent, an algorithm for constructing approximate K-nearest neighbor graphs (K-NN graphs), and RNG Strategy, an algorithm for selecting edges effective for search. This algorithm allows the direct construction of graph-based indexes without ANNS.
  % 実験結果は，提案手法が既存のSOTAと同等の性能を持ちながら，インデックスの構築速度が最も高速であることを示した．例えば，GIST1Mに対する実験では，提案手法の構築時間は，グラフベース手法のSOTAであるNSGの2倍ほど高速であり，これはNN-Descentの構築時間よりも速かった．
  Experimental results demonstrated that the proposed method had the fastest index construction speed, while its search performance is comparable to existing state-of-the-art methods such as NSG. For example, in experiments on the GIST1M dataset, the construction of the proposed method is 2x faster than NSG. Additionally, it was even faster than the construction speed of NN-Descent.
\end{abstract}

\begin{CCSXML}
  <ccs2012>
  <concept>
  <concept_id>10002951.10003227.10003351.10003445</concept_id>
  <concept_desc>Information systems~Nearest-neighbor search</concept_desc>
  <concept_significance>500</concept_significance>
  </concept>
  <concept>
  <concept_id>10010147.10010178.10010224.10010225.10010231</concept_id>
  <concept_desc>Computing methodologies~Visual content-based indexing and retrieval</concept_desc>
  <concept_significance>100</concept_significance>
  </concept>
  </ccs2012>
\end{CCSXML}

\ccsdesc[500]{Information systems~Nearest-neighbor search}
\ccsdesc[100]{Computing methodologies~Visual content-based indexing and retrieval}

\keywords{approximate nearest neighbor search, graph-based index}


\maketitle

\section{Introduction}

Approximate Nearest Neighbor Search (ANNS) is the task of finding the database vector that is closest to a given query vector. It has many applications, from image and text retrieval to machine learning, such as the k-nearest neighbor classification. ANNS has a tradeoff between speed, accuracy, and memory consumption. Typical methods for solving ANNS include hashing-based~\cite{gionis1999similarity,NIPS2015_2823f479}, quantization-based~\cite{jegou2010product,opq}, and tree-based~\cite{muja2009fast, WangWJLZZH14} approaches.

In recent years, many studies have proposed graph-based ANNS methods.
Graph-based methods construct a graph with database vectors as vertices. The edges of the graph connect vectors that are close to each other. The search algorithm finds the approximate nearest neighbor by traversing the graph toward the query.
Graph-based ANNS is a family of methods offering the best accuracy and speed tradeoff for million-scale datasets~\cite{li2019approximate, cvpr23_tutorial_neural_search}. 

The weakness of graph-based methods is their long index construction time. For example, consider index construction on the GIST1M~\cite{jegou2010product} dataset, which contains 1 million vectors of 960 dimensions. Construction of HNSW~\cite{malkov2018efficient}, a representative of graph-based indexes, takes approximately 20 minutes on an AWS c6i.4xlarge instance (16 vCPUs, 32GB memory).
Speeding up the index construction time is important for many cases. One of those cases is when data is updated frequently. Only a few studies~\cite{singh2021freshdiskann} have addressed the issue of efficiently deleting data from a graph-based index. Therefore, when deleting a large amount of data from an index, users must reconstruct the graph from scratch. Fast construction also improves efficiency in finding optimal construction parameters for various datasets.
Despite their importance, there is little research on speeding up index construction.

We aim to accelerate the construction of graph indexes.
We classify conventional graph-based indexes into two types and show why each is slow to construct.
The first type is (1) the refinement-based approach~\cite{FuNSG17, nssg,iwasaki2016pruned,iwasaki2018optimization,li2019approximate}. These methods first construct an initial graph (e.g., an approximate K-NN graph by NN-Descent~\cite{dong2011efficient}). Then they refine the K-NN graph using algorithms such as RNG Strategy~\cite{harwood2016fanng, malkov2018efficient, FuNSG17} to obtain the final graph. The refinement-based approach can construct high-performance graph indexes quickly by using K-NN graphs to narrow candidate edges of the index. However, the construction of the initial K-NN graph takes time.
The second type is (2) the direct approach~\cite{malkov2014approximate, malkov2018efficient,jayaram2019diskann}, which constructs a graph index directly. HNSW~\cite{malkov2018efficient} is a typical method of this type. HNSW sequentially adds new vectors to the index. Here, HNSW finds neighbors of the new data by performing ANNS on the index under construction. HNSW is one of the best-performing graph-based methods. However, constructing a high-performance graph requires accurate ANNS, which is time-consuming because of the tradeoff between ANNS speed and accuracy.

We propose a new graph index construction algorithm, Relative NN-Descent (RNN-Descent). RNN-Descent simultaneously solves the problems of the refinement-based approach and the direct approach. First, RNN-Descent directly constructs the graph-based index to avoid the construction of K-NN graphs, which is a problem of the refinement-based approach. Second, RNN-Descent constructs the graph using the algorithm derived from NN-Descent. This algorithm does not use ANNS, unlike the direct approach.
The technical idea of the proposed method is to execute the construction and improvement of K-NN graphs in the refinement-based approach simultaneously. Specifically, RNN-Descent combines NN-Descent, an algorithm for constructing K-NN graphs, and RNG Strategy, an algorithm for selecting edges valid for ANNS. This idea allows the algorithm to directly construct the index without using ANNS. The neighborhood update algorithm also naturally preserves connectivity, which is important for graph index performance.

Our experiments showed that the proposed method has the equivalent search performance to the conventional methods, whereas its construction is much faster than theirs. For example, on the GIST1M dataset, the construction speed of the proposed method was about twice that of NSG~\cite{FuNSG17}, one of the state-of-the-art methods. Remarkably, the construction speed of the proposed method was even faster than that of the K-NN graph. We also constructed graphs on the SIFT20M~\cite{johnson2019billion} dataset to analyze their performance on large datasets.

Our contributions are as follows:
\begin{itemize}
  \item We tackle the fast construction of graph indexes. Despite its importance, previous researches have yet to address speeding up the construction of graphs.
  \item We propose RNN-Descent, a novel graph index construction algorithm. RNN-Descent simultaneously solves the problems of the conventional graph-based approaches: (1) the refinement-based approach and (2) the direct approach.
  \item Our experiments demonstrate that RNN-Descent is the fastest construction algorithm, and the constructed index has a compatible search performance with conventional methods.
\end{itemize}

\section{Related work}

\subsection{Approximate Nearest Neighbor Search (ANNS)}

We first formulate the nearest neighbor search (NNS) problem. Let $\bm{q} \in \mathbb{R}^{d}$ as query vector, $n \in \mathbb{Z}$ be the number of database vectors, $\mathcal{X} = \{ \bm{x}_1 , \dots, \bm{x}_n \} \subset \mathbb{R}^d$ be the database vectors and $\mathrm{dist}: \mathbb{R}^d \times \mathbb{R}^d \to \mathbb{R}$ be a distance function. Here, NNS is a task that returns the ID of the database vector $\bm{x}_{i^*} \in \mathcal{X}$ closest to $\bm{q}$:
\begin{equation}
    i^* = \argmin_{i \in \{ 1, \dots, n \}} \mathrm{dist}(\bm{q}, \bm{x}_i)
\end{equation}
A naive approach to solving NNS is to compute the distance between all the data and the query. However, the time complexity of the exhaustive search is $O(nd)$, which is slow for large and high-dimensional data.

Numerous studies have addressed approximate NNS (ANNS) as a method for large-scale data. ANNS is a method that improves speed significantly at the cost of a slight loss of accuracy. ANNS methods generally have a trade-off between accuracy, speed, and memory consumption. Therefore the appropriate method depends on the scale of the problem.
ANNS methods include hash-based~\cite{gionis1999similarity, NIPS2015_2823f479}, quantization-based~\cite{jegou2010product, opq}, and tree-based methods~\cite{muja2009fast, WangWJLZZH14}.

\subsection{Graph-Based ANNS}

\begin{algorithm}[tb]
  \caption{\textsc{Search}($G, \bm{q}, L$)}
  \label{alg:graph_search}
  \begin{algorithmic}[1]
      \Require graph $G = (V, E)$, query $\bm{q} \in \mathbb{R}^d, L \in \mathbb{Z}$
      \Ensure approximate nearest neighbor $v^* \in V$
      \State $C \gets$ \textsc{InitializeCandidates}()
      \While{True}
          \State $u \gets$ nearest unvisited point to $\bm{q}$ in $C$
          \State $U \gets \{ v | (u, v) \in E \}$
          \For{$v \in U$}
              \If{$v$ is not visited}
                  \State $C \gets C \cup \{v\}$
              \EndIf
          \EndFor
          \If{$|C| > L$}
              \State $C \gets$ top $L$ nearest points to $\bm{q}$ in $C$
          \EndIf
          \If{$C$ is not updated}
              \State break
          \EndIf
      \EndWhile
      \State \Return nearest point to $\bm{q}$ in $C$
  \end{algorithmic}
\end{algorithm}

This section describes graph-based ANNS.
Graph-based methods have the best trade-off between accuracy and speed for million-scale datasets~\cite{li2019approximate,graph_survey,aumuller2017ann}. Before searching, graph-based methods construct a graph in which vertices represent database vectors. The edges of the graph connect vectors close to each other. The search algorithm finds the approximate nearest neighbor by traversing the constructed graph toward the query.

Algorithm~\ref{alg:graph_search} shows the typical algorithm of graph traversal. Here, $V = \{1, \dots, n\}$ is the graph vertex set, $E \subset V \times V$ is the edge set, and $L \in \mathbb{Z}$ is a hyperparameter. Each $v \in V$ corresponds to a database vector $\bm{x}_v \in \mathbb{R}^d$. First, the algorithm initializes the set of candidate neighborhood points $C \subset V$ (L1). Each step in the while loop takes the unvisited point closest to the query $u$ from $C$ (L3), then adds unvisited vertices of $u$'s neighbors to $C$ (L4-7). Subsequently, the algorithm selects the top $L$ closest points to $\bm{q}$ in $C$ so that the size of $C$ does not exceed $L$ (L8-9). When $C$ is no longer updated, the algorithm terminates the while loop and outputs the nearest neighbor from $C$ (L10-11).

One of the most straightforward graph indexes is the K-nearest neighbor graph (K-NN graph). A K-NN graph has edges from each vertex to the top K vertices closest to it. Since constructing an exact K-NN graph is time-consuming, several methods construct approximate K-NN graphs. 
NN-Descent~\cite{dong2011efficient} is one of the fastest and most accurate methods. NN-Descent improves K-NN graphs step-by-step by dealing with the neighbors of neighbors as candidates for new neighbors. Other approaches include divide-and-conquer~\cite{chen2009fast, WangWZTGL12}, hashing-based~\cite{fu2016efanna}, and sequentially adding vertices by solving ANNS on a subset of the data~\cite{zhao2021approximate}.

Many methods develop different graph structures to improve search performance.
The first approach is to improve the approximate K-NN graph. Because simple K-NN graphs perform poorly as ANNS indexes, some methods refine K-NN graphs to a high-performance graph index. NSG~\cite{FuNSG17} extracts the edge candidates from the K-NN graph, then selects the necessary ones using RNG Strategy. RNG Strategy is a widely used heuristic for selecting edges based on the distance between neighbors. 
Related methods include those that focus on the angles between edges~\cite{li2019approximate, nssg} or adjust the degree of selection by additional parameters~\cite{jayaram2019diskann}.
The second approach is to construct the graph-based index directly. HNSW~\cite{malkov2018efficient} sequentially adds new data to the index. The construction algorithm finds neighbor candidates of the new vertex by solving ANNS for the current graph index.
HNSW adds data sequentially. These methods consider the newly added data as a query vector and solve ANNS for the current graph index to find candidate neighbors for the new data.
While these construction methods are simpler than those via K-NN graphs, it is slower for high-dimensional datasets where ANNS is difficult.

\begin{algorithm}[tb]
  \caption{\textsc{NNDescentJoin}$(G)$}
  \label{alg:nndescent}
  \begin{algorithmic}[1]
      \Require graph $G = (V, E)$
      \Ensure edge candidates $E' \subset V \times V$
      
      \State $E' \gets E$
      \For{$u \in V$}
        \State $U \gets \{v | (u, v) \in E\}$
        \For{all $(v_1, v_2) \in U \times U, v_1 < v_2$}
          \If{at least one flag of $v_1$ or $v_2$ is ``new''}
            \State $E' \gets E \cup \{ (v_1, v_2), (v_2, v_1) \}$
          \EndIf
        \EndFor
        \State set the flag ``old'' for all vertices in $U$
      \EndFor
      \State \Return $E'$
  \end{algorithmic}
\end{algorithm}

\begin{algorithm}[tb]
  \caption{\textsc{RNGStrategy}$(u, U)$}
  \label{alg:rng}
  \begin{algorithmic}[1]
      \Require vertex $u \in V$, neighbor candidates $U \subset V$
      \Ensure selected neighbors $U' \subset U$
      
      \State sort $v \in U$ in ascending order of $\delta(u, v)$
      \State $U' \gets \emptyset$
      \For{$v \in U$}
        \State $f \gets \mathrm{true}$
        \For{$w \in U'$}
          \If{$\delta(u, v) \ge \delta(v, w)$}
            \State $f \gets \mathrm{false}$
            \State break
          \EndIf
        \EndFor
        \If{$f$}
          \State $U' \gets U' \cup \{v\}$
        \EndIf
      \EndFor
      \State \Return $U'$
  \end{algorithmic}
\end{algorithm}

\begin{figure}[t]
  \centering
  \begin{subfigure}{0.32\linewidth}
    \includegraphics[width=\linewidth]{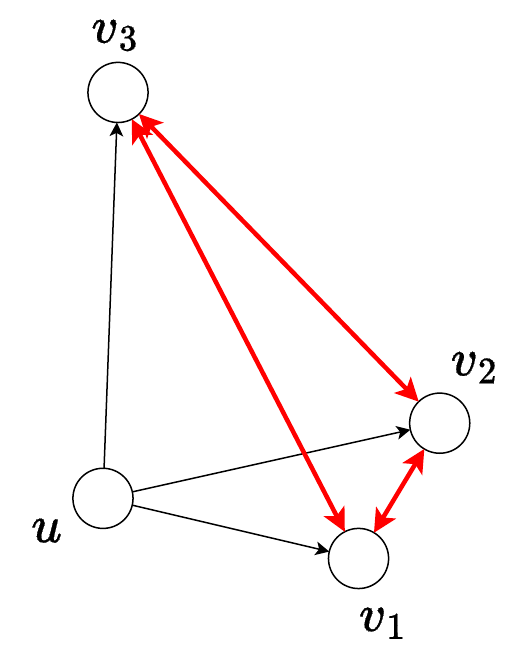}
    \caption{NN-Descent}
    \label{fig:nndescent}
  \end{subfigure}
  \hfill
  \begin{subfigure}{0.32\linewidth}
    \includegraphics[width=\linewidth]{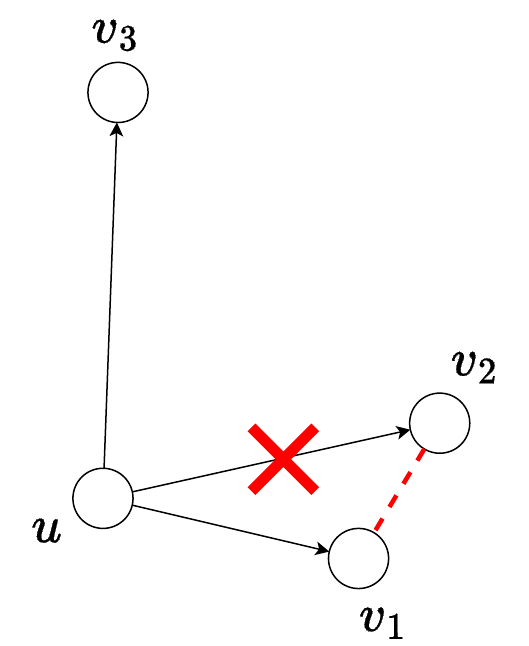}
    \caption{RNG Strategy}
    \label{fig:rng}
  \end{subfigure}
  \hfill
  \begin{subfigure}{0.32\linewidth}
    \includegraphics[width=\linewidth]{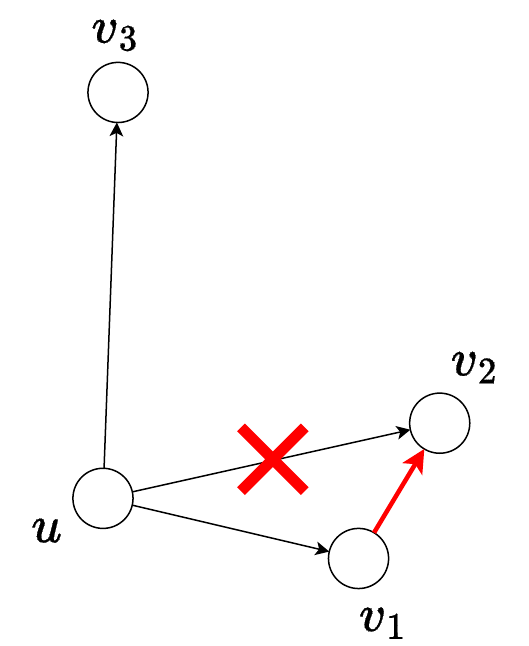}
    \caption{RNN-Descent}
    \label{fig:rnndescent}
  \end{subfigure}
  \caption{
    Comparison of methods for constructing graph indices. Vertices $v_1$, $v_2$, and $v_3$ in the figure are neighbors of vertex $u$. The indexes indicate the order of their distance from $u$.
    (\subref{fig:nndescent}) NN-Descent is a method for constructing approximate K-NN graphs step-by-step by collecting neighbors of neighbors as candidates for new neighbors. For example, vertices $v_1$ and $v_2$ are neighbors of a neighbor via $u$. NN-Descent adds a new bi-directional edge between $v_1$ and $v_2$.
    (\subref{fig:rng}) RNG Strategy is one of the methods to select edges essential for the search. In the figure, RNG Strategy removes the edge to $v_2$ because $v_1$ and $v_2$ are close enough.
    (\subref{fig:rnndescent}) RNN-Descent, the proposed method, combines the features of NN-Descent and RNG Strategy. In the figure, it removes the edge $(u, v_2)$ by RNG Strategy and adds a new edge $(v_1, v_2)$ instead.
  }
  \label{fig:method_overview}
\end{figure}

\section{Preliminary}
\label{sec:Preliminary}

\subsection{NN-Descent}

NN-Descent~\cite{dong2011efficient} is a fast algorithm for constructing approximate K-NN graphs. The basic idea of NN-Descent is that neighbors of neighbors are likely to be neighbors again. NN-Descent first initializes the graph randomly. It then repeats the join step to find new neighbor candidates and the update step to select the neighbors from the candidates.

Figure~\ref{fig:method_overview}(\subref{fig:nndescent}) shows an example of the join operation. For speed, the join operation examines all pairs around an arbitrary vertex instead of checking the neighbors of the neighbors. The vertices $v_1, v_2$, and $v_3$ in Figure~\ref{fig:method_overview}(\subref{fig:nndescent}) are neighbors of a neighbor via $u$. Therefore, the join algorithm adds a bi-directional edge between each vertex pair $(v_i, v_j)$ $(1 \le i < j \le 3)$.

Algorithm~\ref{alg:nndescent} is a pseudo-code for the NN-Descent join operation. The algorithm takes graph $G = (V, E)$ and returns the candidates of new edges. Each loop after Line~2 bidirectionally adds edges for each pair of $u$'s neighbors.
NN-Descent assigns a flag to each neighbor to determine if it is newly added in the most recent step. Each join step adds an edge between neighbor pairs only if either of the flags is ``new'' (L5). This method allows the join step to check each pair of neighbors only once, speeding up the join step. Since the proposed method incorporates the flags, we describe the algorithm in detail again in Section~\ref{sec:UpdateNeighbors}.

\subsection{RNG Strategy}

The RNG Strategy~\cite{harwood2016fanng,malkov2018efficient,FuNSG17} is a method for selecting edges of a graph valid for ANNS. Figure~\ref{fig:method_overview}(\subref{fig:rng}) shows an image of the RNG Strategy. RNG Strategy reduces edges so that any two neighbors $v$ and $w$ of each vertex $u$ satisfy the following inequalities:
\begin{equation}
  \delta(u, v) < \delta(v, w) \land \delta(u, w) < \delta(v, w)
  \label{eq:rng}
\end{equation}
Here, $\delta(u, v)$ is the distance between $u$ and $v$:
\begin{equation}
  \delta(u, v) = \mathrm{dist}(\bm{x}_u, \bm{x}_v)
\end{equation}
For example, vertices $v_1$ and $v_2$ in Figure~\ref{fig:method_overview}(\subref{fig:rng}) do not satisfy $\delta(u, v_2)$ < $\delta(v_2, v_1)$, so the algorithm deletes the edge from $u$ to $v_2$. Intuitively, when $v$ and $w$ are too close to each other, it is likely that an edge already exists between $v$ and $w$. Then, if there is an edge from $u$ to either $v$ or $w$, the other is also reachable from $u$.

Algorithm~\ref{alg:rng} is a pseudocode for RNG Strategy. The input is neighbor candidates $U$ of vertex $u$, and the output is selected neighbors $U'$. First, the algorithm sorts $U$ in ascending order of distance to $u$ (L1). Then, for each vertex $v$ in $U$, the algorithm determines whether output candidates set $U'$ includes $v$ (L3-8). Specifically, for each vertex $w$ already added to $U'$, it checks whether $v$ satisfies the constraint $\delta(u, v) < \delta(v, w)$. If $v$ passes the check, the algorithm adds $v$ to $U'$ (L9-10).

\begin{algorithm}[t]
  \caption{\textsc{UpdateNeighbors}$(G)$}
  \label{alg:update_neighbors}
  \begin{algorithmic}[1]
      \Require graph $G = (V, E)$, vertex $u \in V$
      
      \For{$u \in V$}
        \State $U \gets \{v | (u, v) \in E\}$
        \State sort $v \in U$ in ascending order of $\delta(u, v)$
        \State $U' \gets \emptyset$
        \For{$v \in U$}
          \State $f \gets \mathrm{true}$
          \For{$w \in U'$}
            \If{both flags of $v$ and $w$ are ``old''}
              \State continue
            \EndIf
            \If{$\delta(u, v) \ge \delta(v, w)$}
              \State $f \gets \mathrm{false}$
              \State $E \gets (E \setminus (u, v)) \cup (w, v)$
              \State break
            \EndIf
          \EndFor
          \If{$f$}
            \State $U' \gets U' \cup \{v\}$
          \EndIf
        \EndFor
        \State set the flag ``old'' for all vertices in $U'$
      \EndFor
  \end{algorithmic}
\end{algorithm}

\section{Method}

\subsection{Motivation}

We categorize conventional graph-based ANNS methods into two main approaches: (1) a refinement-based approach and (2) a direct approach. The refinement-based approach first constructs an approximate K-NN graph and then refines it to obtain a final graph-based index. However, constructing the K-NN graph is time-consuming and increases the overall index construction time. On the other hand, the direct approaches construct the graph-based index without going through the K-NN graph by solving ANNS on the index under construction. However, this approach is also slow because the accuracy of the ANNS must be high to construct a graph-based index with good performance.

We propose a new graph construction algorithm, Relative NN-Descent (RNN-Descent), to solve the above problems. RNN-Descent is faster than the refinement-based approach because it does not go through the K-NN graph. In addition, the proposed method can construct a high-performance graph-based index in less time than the direct approach because it constructs the index without ANNS.

The technical core of the proposed method is the combination of two algorithms mentioned in Section~\ref{sec:Preliminary}: NN-Descent and RNG Strategy. Akin to NN-Descent, RNN-Descent constructs a graph-based index by incrementally improving a randomly initialized graph. However, the update algorithm of RNN-Descent simultaneously performs an edge-adding operation derived from the NN-Descent and an edge-removing operation based on the RNG Strategy. This approach allows the proposed method to directly construct the graph without finding neighbor candidates with ANNS. In addition, the proposed update algorithm naturally guarantees graph connectivity, which is essential for search performance.

Section~\ref{sec:UpdateNeighbors} describes the details of the proposed neighbors updating algorithm, and Section~\ref{sec:AddReverseEdges} discusses the reverse edges addition algorithm to avoid suboptimal graphs. Finally, Section~\ref{sec:Overall} describes the overall construct and search algorithm.

\begin{algorithm}[tb]
  \caption{\textsc{AddReverseEdges}$(G, R)$}
  \label{alg:add_reverse_edges}
  \begin{algorithmic}[1]
      \Require graph $G = (V, E)$, $R \in \mathbb{Z}$
      
      \State $E \gets E \cup \{(v, u) | (u, v) \in E\}$
      \State set flags of new neighbors to ``new''
      \For{$v \in V$}
        \State $E_u \gets \{(v, u) | (v, u) \in E\}$
        \State remove top-$R$ shortest edges from $E_u$
        \State $E \gets E \setminus E_u$
      \EndFor

      \For{$v \in V$}
        \State $E_u \gets \{(u, v) | (u, v) \in E\}$
        \State remove top-$R$ shortest edges from $E_u$
        \State $E \gets E \setminus E_u$
      \EndFor
  \end{algorithmic}
\end{algorithm}

\begin{algorithm}[tb]
  \caption{\textsc{RNN-Descent}$(S, R, T_1, T_2)$}
  \label{alg:rnn_descent}
  \begin{algorithmic}[1]
      \Require $S, R, T_1, T_2 \in \mathbb{Z}$
      \Ensure graph $G = (V, E)$
      
      \State $G \gets $ \textsc{RandomGraph}$(S)$
      \State initialize all flags to ``new''
      \For{$t_1 = 1, \dots, T_1$}
        \For{$t_2 = 1, \dots, T_2$}
          \State \textsc{UpdateNeighbors}$(G)$
        \EndFor
        \If{$t_1 \neq T_1$}
          \State \textsc{AddReverseEdges}$(G, R)$
        \EndIf
      \EndFor

      \State \Return $G$
  \end{algorithmic}
\end{algorithm}

\begin{figure*}[t]
  \centering
  \begin{subfigure}{0.33\linewidth}
    \includegraphics[width=\linewidth, scale=0.3]{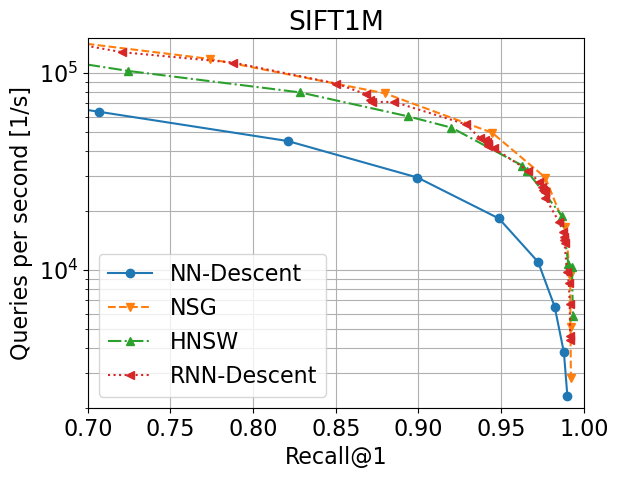}
    \caption{SIFT1M}
    \label{fig:compare_search_sift1m}
  \end{subfigure}
  \hfill
  \begin{subfigure}{0.33\linewidth}
    \includegraphics[width=\linewidth, scale=0.3]{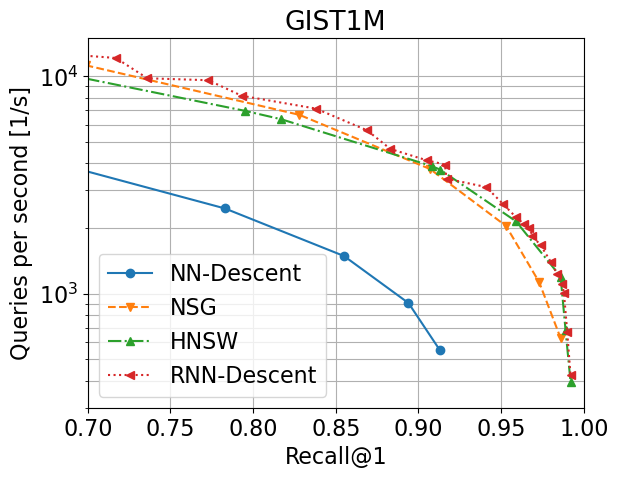}
    \caption{GIST1M}
    \label{fig:compare_search_gist}
  \end{subfigure}
  \hfill
  \begin{subfigure}{0.33\linewidth}
    \includegraphics[width=\linewidth, scale=0.3]{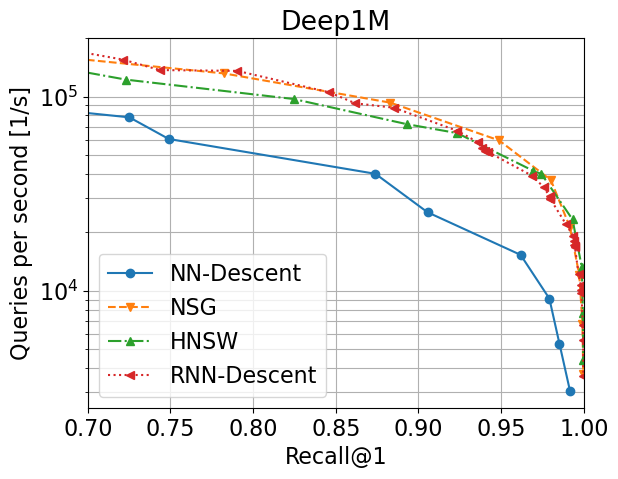}
    \caption{Deep1M}
    \label{fig:compare_search_deep1m}
  \end{subfigure}
  \caption{
  Search performance for graph-based ANNS methods.
  }
  \label{fig:compare_search}
\end{figure*}

\begin{figure*}[t]
  \centering
  \begin{subfigure}{0.33\linewidth}
    \includegraphics[width=\linewidth, scale=0.3]{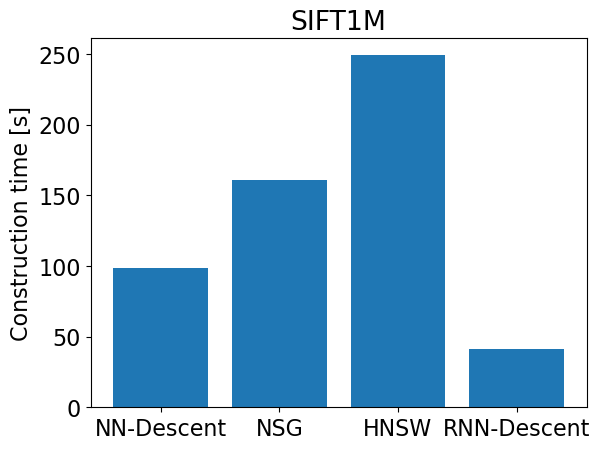}
    \caption{SIFT1M}
    \label{fig:compare_construction_sift1m}
  \end{subfigure}
  \hfill
  \begin{subfigure}{0.33\linewidth}
    \includegraphics[width=\linewidth, scale=0.3]{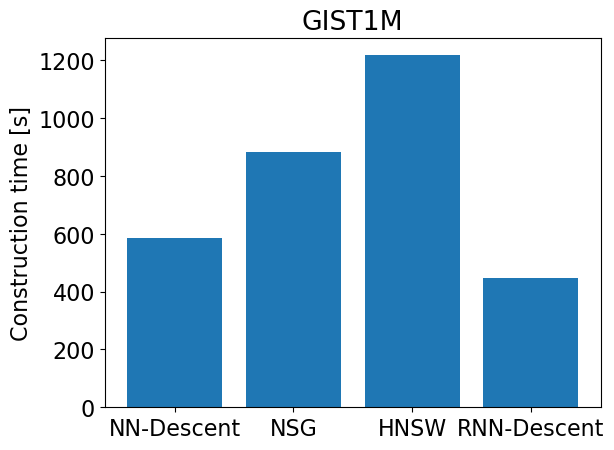}
    \caption{GIST1M}
    \label{fig:compare_construction_gist}
  \end{subfigure}
  \hfill
  \begin{subfigure}{0.33\linewidth}
    \includegraphics[width=\linewidth, scale=0.3]{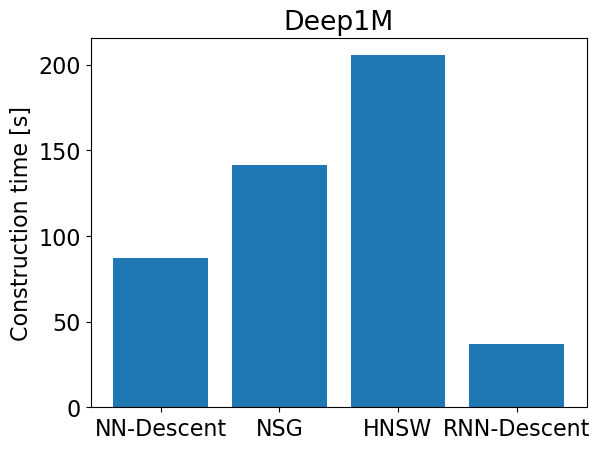}
    \caption{Deep1M}
    \label{fig:compare_construction_deep1m}
  \end{subfigure}
  \caption{
  Construction time for graph-based ANNS methods.
  }
  \label{fig:compare_construction}
\end{figure*}

\subsection{Updating neighbors}
\label{sec:UpdateNeighbors}

This section describes the algorithm for updating the neighbors.
The idea of the proposed method is to simultaneously perform the neighborhood update algorithm of NN-Descent and the edge selection algorithm of the RNG Strategy. 

Figure~\ref{fig:method_overview} shows how the proposed algorithm works. We consider neighbors of a vertex $u$. Normal NN-Descent adds an edge between any two neighborhoods. For example, in Figure~\ref{fig:method_overview}(\subref{fig:nndescent}) NN-Descent adds bidirectional edges between $v_1, v_2$, and $v_3$. However, adding edges between every neighbor pair is useless in constructing a graph-based index. It is because some edges will be eliminated by algorithms such as RNG-Strategy, as the conventional refinement-based approach does.
Therefore, the proposed method uses the RNG Strategy concept to add only the necessary edges. According to RNG Strategy, the edge $(u, v_2)$ in Figure~\ref{fig:method_overview}(\subref{fig:rng}) is not necessary because of the inequality $\delta(u, v_2) > \delta(v_2, v_1)$.
The proposed method removes redundant edges and adds proper edges simultaneously. In Figure~\ref{fig:method_overview}(\subref{fig:rnndescent}), the algorithm removes the edge $(u, v_2)$ and then inserts the edge $(v_2, v_1)$ instead. That is, the proposed method combines NN-Descent and RNG Strategy. Our neighborhood update algorithm also keeps the graph's connectivity, which is important to improve the performance of ANNS. For instance, in Figure~\ref{fig:method_overview}(\subref{fig:rnndescent}), $v_2$ is reachable from $u$ before and after the update algorithm.

Algorithm~\ref{alg:update_neighbors} is a pseudo code for a neighborhood update algorithm. The algorithm takes graph $G=(V, E)$ as input and updates the edge set $E \subset V \times V$.
Most of the algorithm is the same as the RNG Strategy. One of the differences is in Line~11. If $u$'s neighbor $v$ does not satisfy the inequality for some selected vertex $w$, the algorithm removes edge $(u, v)$ and inserts the edge $(v, w)$ instead.
The algorithm adds no edges if the edge $(w, v)$ already exists. However, $w$ is still reachable from $u$ in this case.

Another difference is the introduction of a flag to determine if each neighbor is newly added in the last iteration. This technique derives from the original NN-Descent. If both flags of vertices $v$ and $w$ are ``old,'' the algorithm skips to calculate the distance between them (L5-6). It is because if $v$ and $w$ are old neighbors, the algorithm has already checked whether $v$ and $w$ satisfy the RNG inequality. At the end of the iteration, the algorithm sets the flag of all neighbors in $U'$ to ``old'' (L15).

\subsection{Adding reverse edges}
\label{sec:AddReverseEdges}

The problem with the update algorithm in Section~\ref{sec:UpdateNeighbors} is that the constructed graph will likely fall into a local optimum with low performance. Here, a local optimum means that all neighbors satisfy the conditions of the RNG Strategy, and the update algorithm will not update edges anymore. Suboptimal graphs have long average edge distances, leading to poor search performance.

Our solution is to add reverse edges to the suboptimal graph. Adding a new edge that does not satisfy Eq~\ref{eq:rng} allows the update algorithm to restart. Also, since the graph's reverse edges are likely shorter than randomly chosen edges, they are suitable for converging the graph to a better solution.

Algorithm~\ref{alg:add_reverse_edges} is a pseudo-code for adding the reverse edges. Here, the input $R \in \mathbb{Z}$ is a parameter that controls the number of reduced edges.
First, the algorithm adds the reverse edges to the current edge set $E$ (L1). Then, it removes some long edges from $E$ to prevent the number of edges from increasing too much. Specifically, the algorithm reduces the edges so that the in-degree (L3-5) and out-degree are less than or equal to $R$ (L6-8).

\begin{figure*}[t]
  \centering
  \begin{subfigure}{0.33\linewidth}
    \includegraphics[width=\linewidth, scale=0.3]{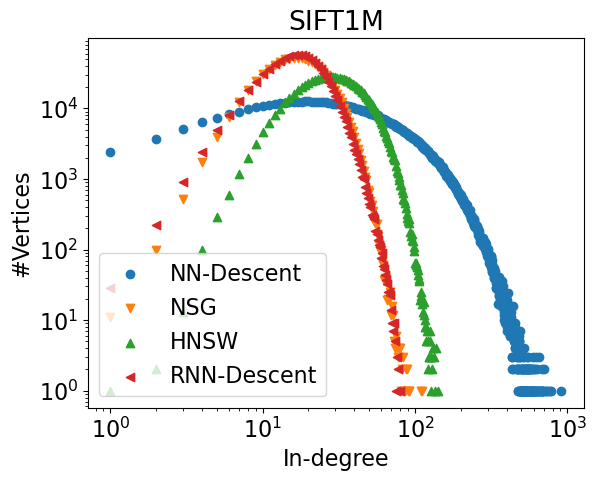}
    \caption{SIFT1M}
    \label{fig:indegree_sift}
  \end{subfigure}
  \hfill
  \begin{subfigure}{0.33\linewidth}
    \includegraphics[width=\linewidth, scale=0.3]{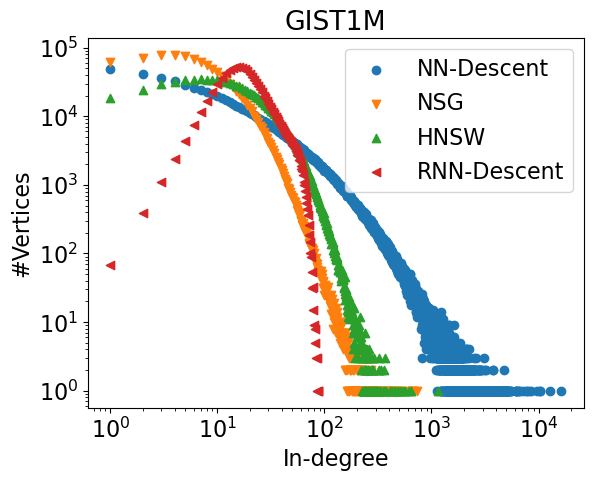}
    \caption{GIST1M}
    \label{fig:indegree_gist}
  \end{subfigure}
  \hfill
  \begin{subfigure}{0.33\linewidth}
    \includegraphics[width=\linewidth, scale=0.3]{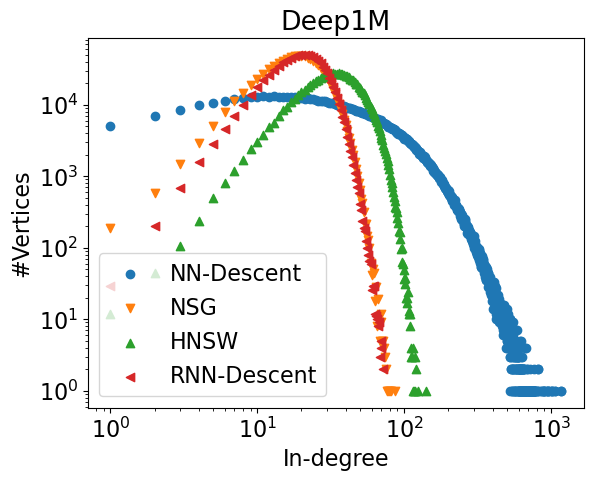}
    \caption{Deep1M}
    \label{fig:indegree_deep1m}
  \end{subfigure}
  \caption{
  In-degree distribution.
  }
  \label{fig:indegree}
\end{figure*}

\begin{figure*}[t]
  \centering
  \begin{subfigure}{0.33\linewidth}
    \includegraphics[width=\linewidth, scale=0.3]{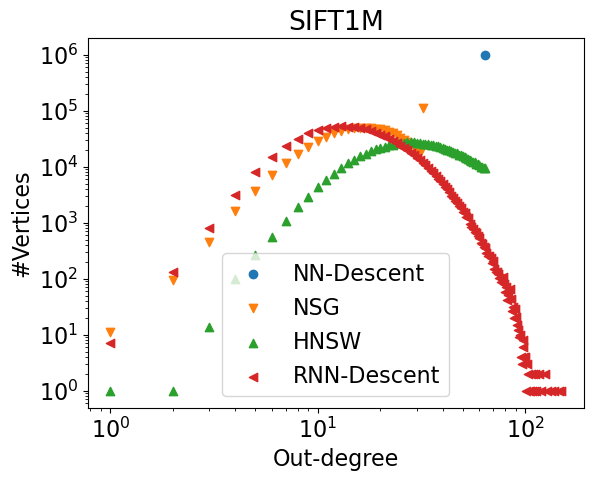}
    \caption{SIFT1M}
    \label{fig:outdegree_sift}
  \end{subfigure}
  \hfill
  \begin{subfigure}{0.33\linewidth}
    \includegraphics[width=\linewidth, scale=0.3]{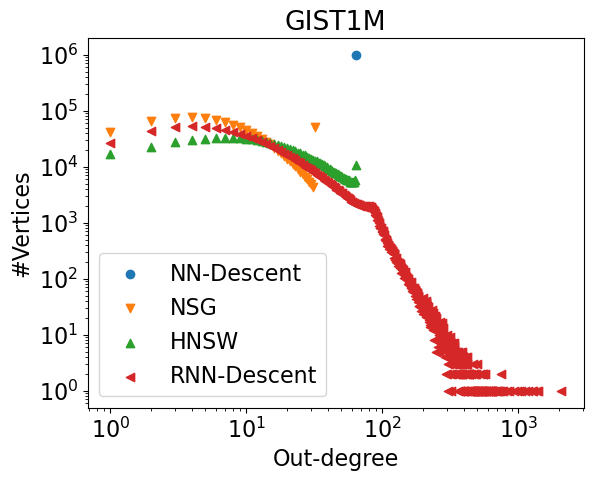}
    \caption{GIST1M}
    \label{fig:outdegree_gist}
  \end{subfigure}
  \hfill
  \begin{subfigure}{0.33\linewidth}
    \includegraphics[width=\linewidth, scale=0.3]{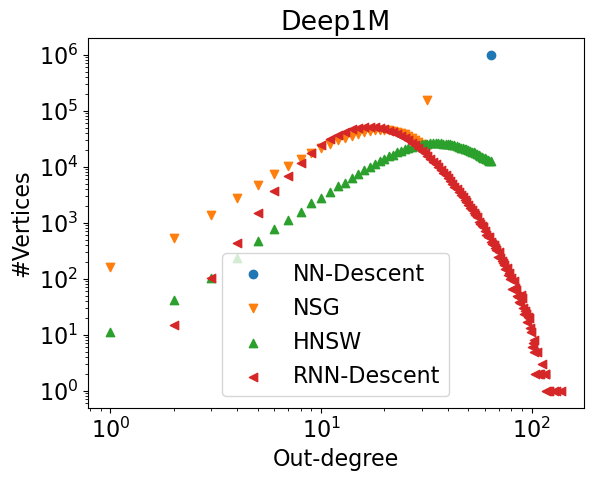}
    \caption{Deep1M}
    \label{fig:outdegree_deep1m}
  \end{subfigure}
  \caption{
  Out-degree distribution.
  }
  \label{fig:outdegree}
\end{figure*}

\subsection{Overall algorithm}
\label{sec:Overall}

Algorithm~\ref{alg:rnn_descent} is the pseudo-code for the entire RNN-Descent algorithm.
First, the algorithm initializes the graph randomly (L1). Here, $S \in \mathbb{Z}$ is the out-degree of the initial graph. Each step in the subsequent loop improves the initialized graph $G$ step-by-step. The algorithm in Section~\ref{sec:UpdateNeighbors} updates the edge set $T_2$ times (L4-5). Then, the algorithm in Section~\ref{sec:AddReverseEdges} adds reverse edges to prevent the graph from falling suboptimal (L6-7). After repeating this sequence of steps $T_1$ times, the algorithm outputs the final graph (L8). 

Finally, we describe the search algorithm for the constructed indexes. RNN-Descent does not limit the out-degree of the constructed graph. 
Instead, the search algorithm limits the number of degrees. We replace L4 in Algorithm~\ref{alg:graph_search} as follows:
\begin{equation}
    U \gets \mathrm{top}\ K\ \mathrm{nearest}\ \mathrm{points}\ \mathrm{to}\ u\ \mathrm{in}\ \{ v | (u, v) \in E \}
\end{equation}
Here, $K \in \mathbb{Z}$ is a parameter that determines the maximum out-degree. This algorithm allows users to change the maximum out-degree without reconstructing the graph.
The optimal degree depends on the dataset, but it is difficult to know this before the construction. In contrast, the proposed method can dynamically determine the optimal out-degree during the search.

\begin{figure*}[t]
  \centering
  \begin{subfigure}{0.33\linewidth}
    \includegraphics[width=\linewidth, scale=0.3]{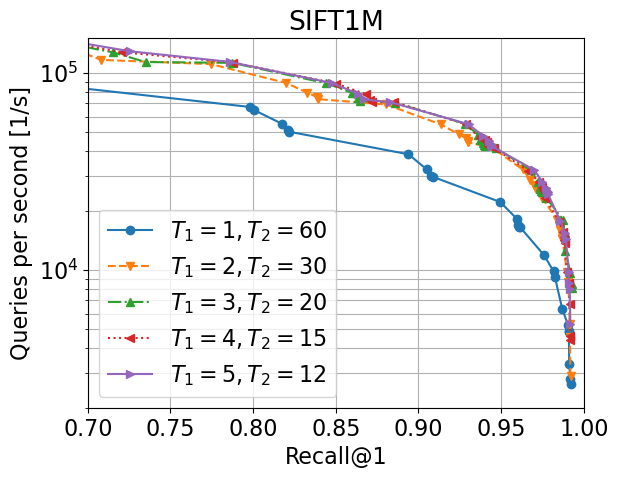}
    \caption{SIFT1M}
    \label{fig:ablation_t_search_sift1m}
  \end{subfigure}
  \hfill
  \begin{subfigure}{0.33\linewidth}
    \includegraphics[width=\linewidth, scale=0.3]{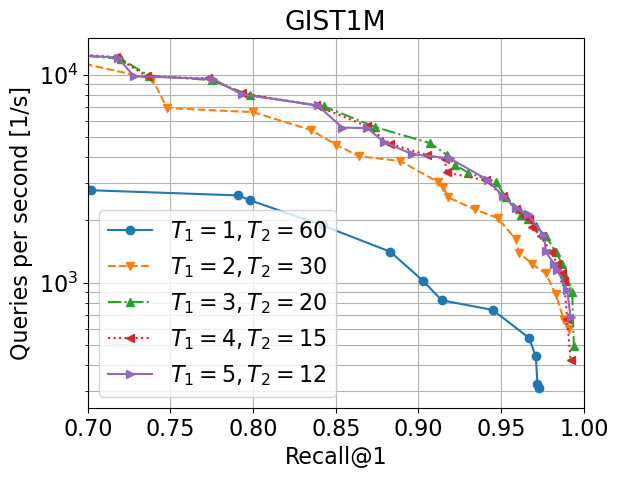}
    \caption{GIST1M}
    \label{fig:ablation_t_search_gist}
  \end{subfigure}
  \hfill
  \begin{subfigure}{0.33\linewidth}
    \includegraphics[width=\linewidth, scale=0.3]{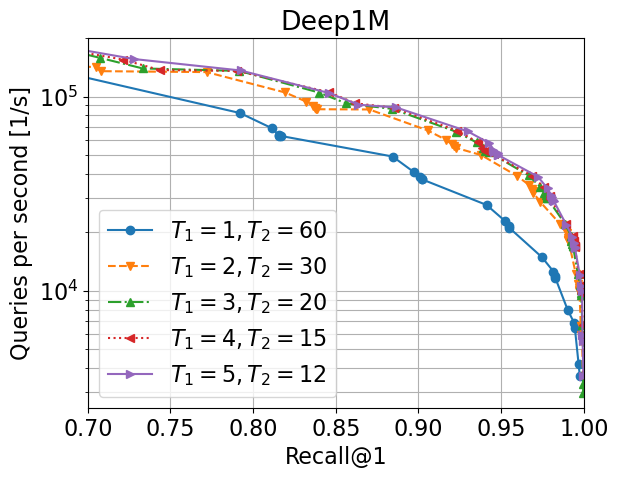}
    \caption{Deep1M}
    \label{fig:ablation_t_search_deep1m}
  \end{subfigure}
  \caption{
  Search performance for various $(T_1, T_2)$.
  }
  \label{fig:ablation_t_search}
\end{figure*}

\begin{figure*}[t]
  \centering
  \begin{subfigure}{0.33\linewidth}
    \includegraphics[width=\linewidth, scale=0.3]{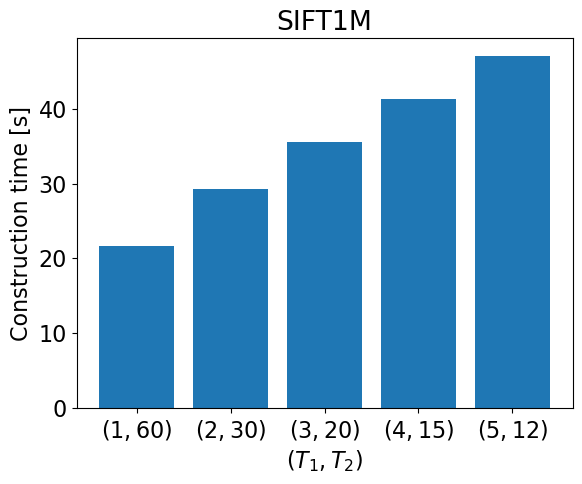}
    \caption{SIFT1M}
    \label{fig:ablation_t_construction_sift1m}
  \end{subfigure}
  \hfill
  \begin{subfigure}{0.33\linewidth}
    \includegraphics[width=\linewidth, scale=0.3]{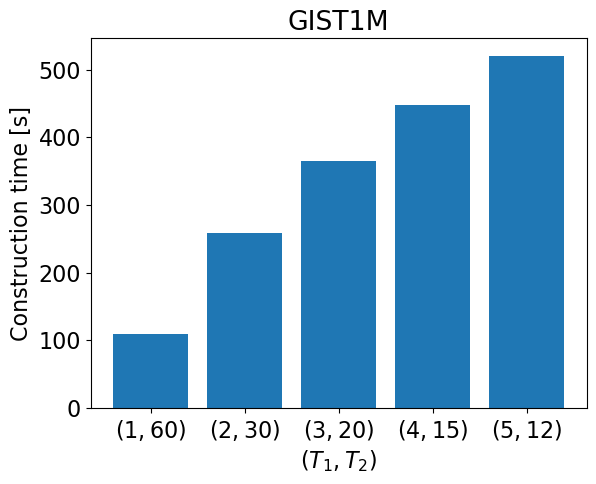}
    \caption{GIST1M}
    \label{fig:ablation_t_construction_gist}
  \end{subfigure}
  \hfill
  \begin{subfigure}{0.33\linewidth}
    \includegraphics[width=\linewidth, scale=0.3]{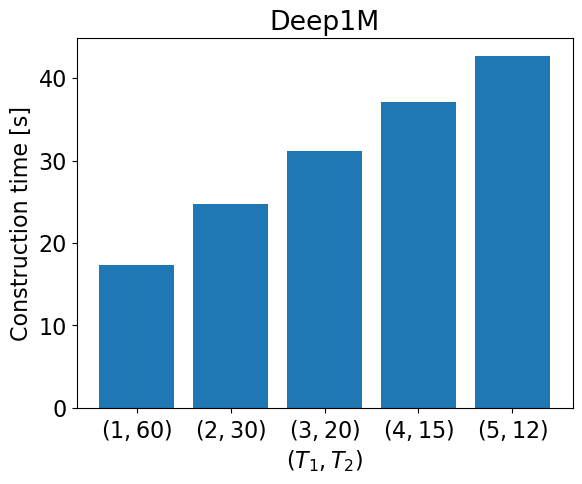}
    \caption{Deep1M}
    \label{fig:ablation_t_construction_deep1m}
  \end{subfigure}
  \caption{
  Construction time for various $(T_1, T_2)$.
  }
  \label{fig:ablation_t_construction}
\end{figure*}

\begin{table}[t]
  \caption{The properties of datasets.}
  \label{tab:datasets}
  \begin{tabular}{cccc}
    \toprule
    Dataset & Dimensions & \#Bases & \#Queries \\
    \midrule
    SIFT1M~\cite{jegou2010product} & 128 & 1,000,000 & 10,000 \\
    GIST1M~\cite{jegou2010product} & 960 & 1,000,000 & 1,000 \\
    Deep1M~\cite{babenko2016efficient} & 96  & 1,000,000 & 10,000 \\
    SIFT20M~\cite{johnson2019billion} & 128  & 20,000,000 & 10,000 \\
    \bottomrule
  \end{tabular}
\end{table}

\section{Experiments}

\subsection{Settings}
\label{sec:settings}

This experiment measures the construction time and search performance of each graph-based method. The search accuracy metric is Recall@1 (R@1), where R@1 is the percentage of queries that find the correct nearest neighbor. We also use queries per second (QPS) to measure search speed. In general, ANNS methods have parameters specified during a search, which control the balance between speed and accuracy. We evaluate each method by plotting R@1 and QPS on a plane while varying the search parameters.

We compared the following algorithms:
\begin{itemize}
  \item NN-Descent~\cite{dong2011efficient}: An approximate K-NN graph. We set $K=64, S=10, L=114, R=100, \mathrm{iter}=10$.
  \item NSG~\cite{FuNSG17}: The SOTA of refinement-based approach. NSG uses NN-Descent to construct a K-NN graph. We set $R=32, L=64, C=132$. The parameters of NN-Descent are same as above.
  \item HNSW~\cite{malkov2018efficient}: The SOTA of the direct approach. We set $M=32, \mathrm{efC}=500$.
  \item RNN-Descent: The proposed method. We set $S=20, R=96, T_1=4, T_2=15$.
\end{itemize}

We used the Faiss implementation for comparison. We conducted the experiments on an AWS c6i.4xlarge instance (16 vCPUs, 32 GB memory). We set the number of threads to 16. Table~\ref{tab:datasets} summarizes the properties of the datasets we used.

\subsection{Comparison to other methods}

Figure~\ref{fig:compare_search} and~\ref{fig:compare_construction} shows each graph-based method's search performance and construction time. Note that we plot only Pareto-optimal points. 
The search performance of the proposed method was comparable to the existing SOTA method. On the other hand, the construction time was the shortest among all methods.
We emphasize that the construction speed of the proposed method was faster than that of NN-Descent. This result means that existing methods of the refinement-based approach cannot construct the index faster than the proposed method, at least as long as it uses NN-Descent. On the other hand, HNSW belongs to the direct approach and may be faster than the K-NN graph. However, experimental results show that HNSW has the slowest construction time among all the methods.

\subsection{Degree distribution}
\label{sec:DegreeDistribution}

\begin{figure*}[t]
  \centering
  \begin{subfigure}{0.33\linewidth}
    \includegraphics[width=\linewidth, scale=0.3]{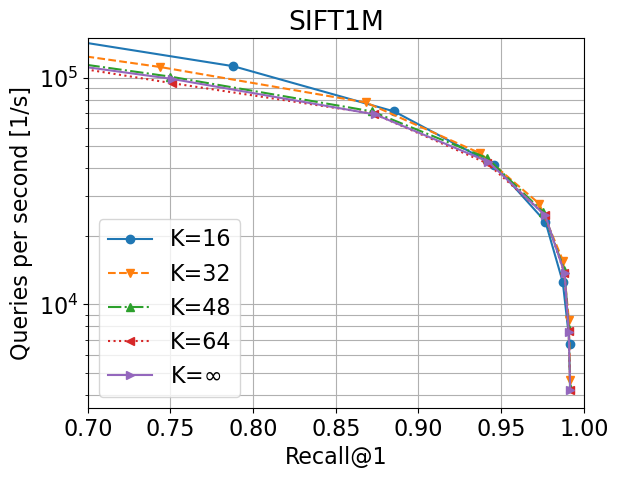}
    \caption{SIFT1M}
    \label{fig:ablation_k_search_sift1m}
  \end{subfigure}
  \hfill
  \begin{subfigure}{0.33\linewidth}
    \includegraphics[width=\linewidth, scale=0.3]{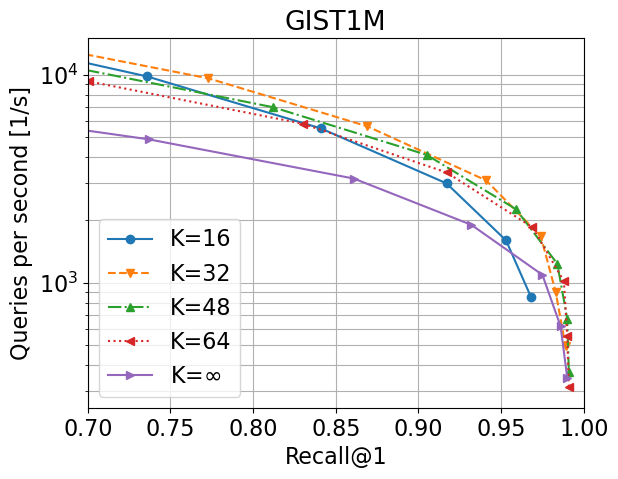}
    \caption{GIST1M}
    \label{fig:ablation_k_search_gist}
  \end{subfigure}
  \hfill
  \begin{subfigure}{0.33\linewidth}
    \includegraphics[width=\linewidth, scale=0.3]{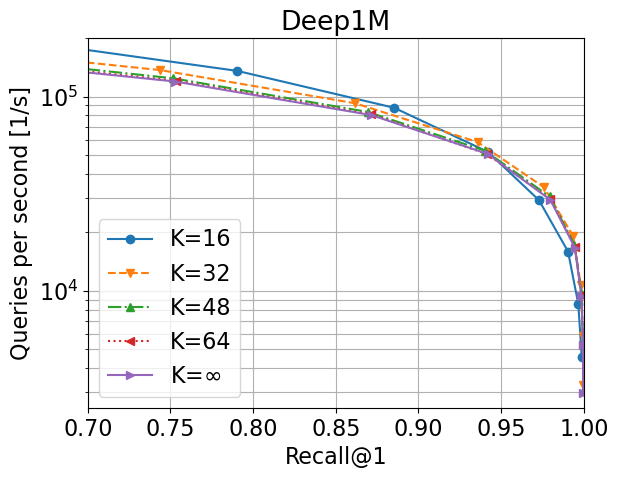}
    \caption{Deep1M}
    \label{fig:ablation_k_search_deep1m}
  \end{subfigure}
  \caption{
  Search performance for various $K$.
  }
  \label{fig:ablation_k_search}
\end{figure*}

Figures~\ref{fig:indegree} and~\ref{fig:outdegree} show the distributions of the in-degree and out-degree of the graphs constructed by each method.
The proposed method limits the in-degree of the graph to be less than or equal to $R$. However, the average degree was much smaller than $R$, around 20, and this value was about the same as existing methods. The proposed method has comparable memory efficiency to the existing methods because the memory consumption of the index is proportional to the average degree of the graph. 

The degree distributions for SIFT1M and Deep1M were similar to those of NSG. Although the proposed method does not limit the out-degree, the maximum out-degree was around 150. These results demonstrate that the proposed method can automatically adjust the out-degree for relatively simple datasets. On the other hand, for the GIST1M dataset, some vertices have a very large out-degree. These vertices slow the search because the search algorithm checks many neighbors when it visits them. Therefore, the proposed method dynamically limits the out-degree during the search to avoid checking too many neighbors. In addition, the input degree of the proposed method has more concentrated peaks than the input degree of other methods.

\subsection{Ablation study}

\paragraph*{Adding reverse edges.}
Figures~\ref{fig:ablation_t_search} and~\ref{fig:ablation_t_construction} show the change in search performance and construction speed when $T_1$ and $T_2$ are changed. We keep the total number of iterations $T_1 T_2$ constant. We also set $S=20$ and $R=96$ for all experiments.
The case $T_1 = 1$ is where the algorithm adds no reverse edges. In this case, the search performance was the lowest of all settings. This result indicates that adding reverse edges is effective in improving search performance.
As $T_1$ increases, the search performance improves while the construction time increases. This result shows that $T_1$ controls search performance and construction time trade-off.

\paragraph*{Limitation on out-degree}

Figure~\ref{fig:ablation_k_search} shows the change in search performance for different $K$. We set $K = 16, 32, 48, 64, 96, and\ \infty$.

We observe the best $K$ is different whether R@1 was greater than approximately 0.95. First, we examine the case when R@1 is less than 0.95. The optimal $K$ was 16 for the SIFT1M and Deep1M datasets and 32 for GIST1M. Next, we observe the case when R@1 $>$ 0.95. For SIFT1M and Deep1M, the search performance was similar if $K \ge 32$. While for GIST1M, the best $K$ were 48 and 64. These results indicate that $K$ should be small when speed is a priority, and $K$ should be large when accuracy is a priority. 
In addition, for GIST1M, setting $K$ to $\infty$ resulted in significant performance degradation due to the significantly large out-degree, as seen in Section~\ref{sec:DegreeDistribution}. By setting $K$ appropriately, we can avoid the problem of poor search performance.
We emphasize that the user can know the optimal $K$ after index construction. Thus, the proposed method is more robust to changes in the dataset than the conventional methods that require setting the maximum out-degree before construction.

\subsection{Experiment for large datasets}

\begin{figure}[t]
  \centering
  \begin{subfigure}{0.49\linewidth}
    \includegraphics[width=\linewidth]{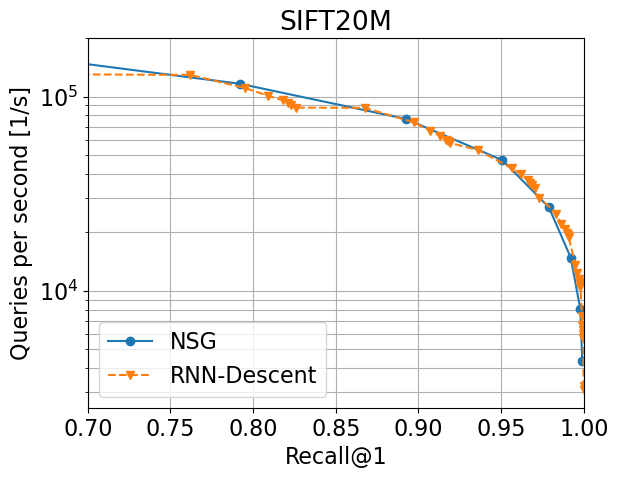}
    \caption{Search performance}
    \label{fig:large_search}
  \end{subfigure}
  \begin{subfigure}{0.49\linewidth}
    \includegraphics[width=\linewidth]{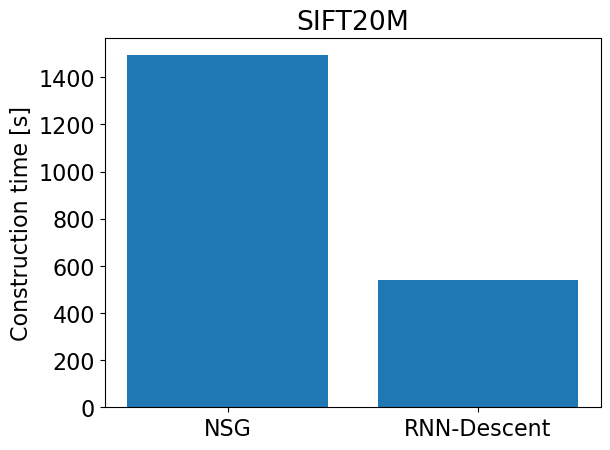}
    \caption{Construction time}
    \label{fig:large_construction}
  \end{subfigure} \\
  \begin{subfigure}{0.49\linewidth}
    \includegraphics[width=\linewidth]{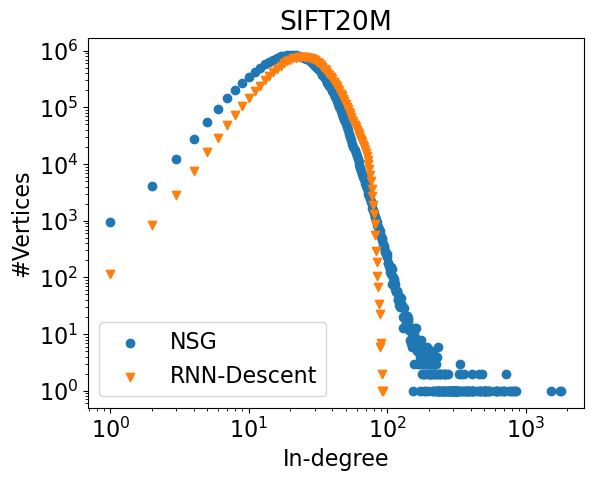}
    \caption{In-degree distribution}
    \label{fig:large_indegree}
  \end{subfigure}
  \begin{subfigure}{0.49\linewidth}
    \includegraphics[width=\linewidth]{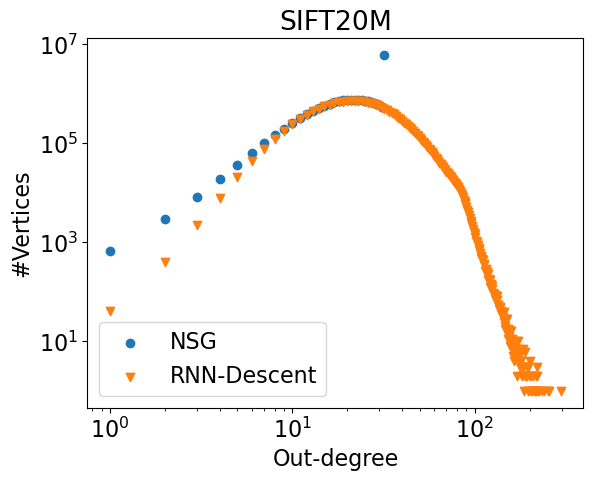}
    \caption{Out-degree distribution}
    \label{fig:large_outdegree}
  \end{subfigure}
  \caption{Experiment for SIFT20M dataset.}
  \label{fig:large}
\end{figure}

We experimented with index construction on a large dataset. We used the SIFT20M dataset, the first 20 million database vectors extracted from the beginning of the SIFT1B~\cite{johnson2019billion} dataset (1 billion vectors, 128 dimensions). We conducted the experiments on an AWS c6i.12xlarge (48 vCPUs, 96GB memory) instance. We set the number of threads to 48.

Figure~\ref{fig:large} shows the search performance and construction time on the SIFT20M dataset. We compared the proposed method to NSG. The parameters of each method are equal to those described in Section~\ref{sec:settings}. First, the proposed method is more than twice as fast as NSG in index construction. On the other hand, the proposed method has comparable search performance to NSG.

\section{Conclusion}

This paper proposes RNN-Descent, a new graph-based ANNS index construction algorithm. 
RNN-Descent combine NN-Descent and RNG Strategy. It simultaneously adds edges based on RNN-Descent and removes edges based on RNG Strategy.
Experimental results show that RNN-Descent significantly accelerates index construction while maintaining performance comparable to existing SOTA methods. For example, experiments on the GIST1M dataset show that RNN-Descent constructs the index approximately twice as fast as NSG.
Our source code is publicly available on \url{https://github.com/mti-lab/rnn-descent}.

\paragraph{Acknowledgement}
This work was supported by JST AIP Acceleration Research JPMJCR23U2, Japan.

\bibliographystyle{ACM-Reference-Format}
\balance
\bibliography{references}

\clearpage
\newcommand\beginsupplement{%
        \setcounter{table}{0}
        \renewcommand{\thetable}{\Alph{table}}%
        \setcounter{figure}{0}
        \renewcommand{\thefigure}{\Alph{figure}}%
     }
\beginsupplement
\appendix

\section{Implementation}

This section describes the details of the implementation. We implemented the program in C++17. The compiler was GCC 11.3.0. We built the source code using CMake\footnote{https://cmake.org}, setting the build type to release. The BLAS used for Faiss~\cite{johnson2019billion} was OpenBLAS\footnote{https://www.openblas.net} 0.3.20.
We used an AWS c6i.4xlarge instance with x86\_64 architecture for our experiment.

\section{Detailed analysis for out-degree}

Table~\ref{tab:degree_comp} summarizes the average out-degree (AOD) for each method. Section 5.4 discusses that the proposed method could optimize the search performance by appropriately setting $K$. Therefore, in addition to the usual AOD, we computed the AOD when we restrict each vertex's out-degree to less than or equal to $K$. Also, HNSW constructs a hierarchical graph, but we only computed the AOD of the bottom graph for simplicity.

First, we check the results for SIFT1M and Deep1M. For these datasets, the optimal $K$ was 16 when R@1 $<$ 0.95 and 32 when R@1 $>$ larger than 0.95, as seen in Section 5.4. In both cases, the AOD of the proposed method was the smallest among all methods. In other words, by reducing unnecessary edges from the index, the proposed method achieves the best memory efficiency without lowering the search performance.

Next, we consider the results for GIST1M. The optimal $K$ for the proposed method is 32 when R@1 $<$ 0.95 and 48 or 64 when R@1 $>$ 0.95. As in the SIFT1M case, the proposed method can reduce the AOD by removing edges, although it is not as low as NSG. For example, with $K$=48, the graph consumes about 18\% less memory than a normal index.

\begin{table}[t]
  \caption{
      Average out-degree (AOD) for each method.
  }
  \label{tab:degree_comp}
  \begin{tabular}{lll}
      \toprule
      Dataset & Method & AOD \\
      \midrule
      \multirow{8}{*}{SIFT1M~\cite{jegou2010product}} & NN-Descent~\cite{dong2011efficient} & 64.0 \\
       & HNSW~\cite{malkov2018efficient} & 35.2 \\
       & NSG~\cite{FuNSG17} & 20.06 \\
       & RNN-Descent ($K$=16) & \textbf{13.93} \\
       & RNN-Descent ($K$=32) & 18.38 \\
       & RNN-Descent ($K$=48) & 19.17 \\
       & RNN-Descent ($K$=64) & 19.31 \\
       & RNN-Descent ($K$=$\infty$) & 19.33 \\
      \midrule
      \multirow{8}{*}{GIST1M~\cite{jegou2010product}} & NN-Descent~\cite{dong2011efficient} & 64.0 \\
       & HNSW~\cite{malkov2018efficient} & 22.38 \\
       & NSG~\cite{FuNSG17} & \textbf{10.63} \\
       & RNN-Descent ($K$=16) & 10.97 \\
       & RNN-Descent ($K$=32) & 15.65 \\
       & RNN-Descent ($K$=48) & 18.19 \\
       & RNN-Descent ($K$=64) & 19.74 \\
       & RNN-Descent ($K$=$\infty$) & 22.13 \\
      \midrule
      \multirow{8}{*}{Deep1M~\cite{babenko2016efficient}} & NN-Descent~\cite{dong2011efficient} & 64.0 \\
       & HNSW~\cite{malkov2018efficient} & 38.59 \\
       & NSG~\cite{FuNSG17} & 21.41 \\
       & RNN-Descent ($K$=16) & \textbf{14.94} \\
       & RNN-Descent ($K$=32) & 20.79 \\
       & RNN-Descent ($K$=48) & 21.83 \\
       & RNN-Descent ($K$=64) & 22.02 \\
       & RNN-Descent ($K$=$\infty$) & 22.06 \\
      \bottomrule
  \end{tabular}
\end{table}

\end{document}